\documentclass[aps,preprint,amssymb,12pt,floatfix]{revtex4}
\usepackage{subfigure}
\usepackage{graphicx}
\usepackage{rotating}
\usepackage{color}
 \usepackage{amsmath,amsthm}
\usepackage{epstopdf}
\setcitestyle{super}
\begin{document}

\title{Folding PDZ2 domain using the Molecular Transfer Model}
\author{Zhenxing Liu$^{1}$, Govardhan Reddy$^{2}$, and D. Thirumalai$^{3,4}$}
\affiliation{$^1$Department of Physics, Beijing Normal University, Beijing, China 100875 \\
$^2${Solid State and Structural Chemistry Unit, Indian Institute of Science, Bangalore, Karnataka, India 560012}\\
$^3$Biophysics Program, Institute For Physical Science and Technology, University of Maryland, College Park, Maryland 20742\\
$^4$Department of Chemistry and Biochemistry, University of Maryland, College Park,
Maryland 20742}

\date{\today}
\begin{abstract}
A major challenge in molecular simulations is to describe denaturant-dependent folding of proteins order to make direct comparisons with {\it in vitro} experiments. We use the molecular transfer model (MTM), which is currently the only method that accomplishes this goal albeit phenomenologically, to quantitatively describe urea-dependent folding of PDZ domain, which plays a significant role in molecular recognition and signaling. Experiments show that urea-dependent unfolding rates of the PDZ2 domain exhibit a downward curvature at high urea concentrations ($[C]$s), which has been interpreted by invoking the presence of a sparsely populated high energy intermediate.  Simulations using the MTM and a coarse-grained Self-Organized Polymer (SOP) representation of PDZ2 are used to show that the intermediate ($I_{EQ}$), which  has some native-like character, is present in equilibrium both in the presence and absence of urea. The free energy profiles as a function of the structural overlap order parameter show that there are two barriers separating the folded and unfolded states. Structures of the transition state ensembles, ($TSE1$ separating the unfolded and $I_{EQ}$ and $TSE2$ separating $I_{EQ}$ and the native state), determined using the $P_{fold}$ method, show that $TSE1$ is greatly expanded while $TSE2$ is compact and native-like. Folding trajectories reveal that PDZ2 folds by parallel routes. In one pathway folding occurs exclusively through $I_1$, which resembles $I_{EQ}$. In a fraction of trajectories, constituting the second pathway, folding occurs through a combination of $I_{1}$ and a kinetic intermediate. We  establish that the radius of gyration ($R_g^{U}$) of the unfolded state is more compact (by $\sim$ 9\%) under native conditions. Theory and simulations show that the decrease in $R_g^{U}$ occurs on the time scale on the order of utmost $\sim$ 20 $\mu s$. The modest decrease in $R_g^{U}$ and the rapid collapse suggest  that high spatial and temporal resolution, currently beyond the scope of most small angle X-ray scattering experiments,  are needed to detect compaction in finite-sized proteins. The present work further establishes that MTM is efficacious in producing nearly quantitative predictions for folding of proteins under conditions used to carry out experiments.       
\end{abstract}
\maketitle

\section*{\Large{Introduction}}

The Molecular Transfer Model (MTM) \cite{O'Brien08PNAS},
based on  the statistical mechanical theory of liquid mixtures  \cite{Liu12JPCB},
is currently the only available computational method that  predicts the outcomes of
experiments and provides the  structural basis of folding as a function of denaturants \cite{Reddy12PNAS,Reddy15JPCB} and pH \cite{OBrien12JACS}.
Using the MTM in conjunction with coarse-grained (CG) representation of polypeptide chains
we have made quantitative predictions of the folding thermodynamics and kinetics as a function of
denaturants for a number of small (protein L, cold shock protein, srcSH3 domain, and Ubiquitin) \cite{O'Brien08PNAS,Liu12JPCB,Reddy15JPCB,Chen14PCCP},
 and GFP, a large single domain proteins \cite{Reddy12PNAS}.  Because the effects of denaturants are
taken into account naturally within the MTM framework \cite{Liu12JPCB},  albeit phenomenologically
it has been possible to obtain chevron plots for src SH3  domain producing quantitative agreement
with experiments for the slopes of folding and unfolding arms \cite{Liu11PNAS}. Although MTM can be be implemented in conjunction with atomically detailed simulations we have so far used CG models for proteins. The virtue of CG models \cite{Tozzini05COSB,Hyeon11NatComm,Elber11inbook,Whitford12RepProgPhys} is that they can be used to obtain both
the thermodynamic and kinetic properties over a a wide range of external conditions,
thus allowing us to compare with experiments directly \cite{Thirumalai13COSB}.
These studies illustrate that simulations based on the MTM provide concrete predictions for {\it in vitro}  experiments enabling us to  go beyond generic ideas  used to understand  protein  folding  \cite{Kim90AnnRevBiochem,Wolynes95Science,Dill97NSB,Daggett03TIBS,Onuchic04COSB,Dill08ARB,Schuler08COSB,Shakhnovich06ChemRev,Thirumalai10ARB,Dill12Science,Piana14COSB}.
Here, we investigate the folding mechanism of PDZ2 domain using CG simulations within the theoretical framework of the MTM.

PDZ domains are a large family of globular proteins that mediate
protein-protein interactions and play an important role in molecular
recognition
\cite{Harris01JCellSci,Kim04NatRevNeurosci,Jemth07Biochem}. These
proteins generally consist of $80-100$ amino acids.   PDZ2 domain has six $\beta$ strands and two $\alpha$
helices (Figure 1a). The folding mechanism of PDZ2, with 94 residues, has been studied experimentally
\cite{Gianni05PEDS} using classical chemical kinetics methods. The key findings in these experiments are: (i) In urea-induced equilibrium denaturation experiments, the observed
transition is cooperative, which is well described by an
apparent two-state model \cite{Gianni05PEDS,Sicorello09BJ}. (ii) In a majority of cases proteins that fold thermodynamically in a two-state manner also exhibit a similar behavior kinetically in ensemble experiments. However, urea-dependent unfolding rates exhibit a downward curvature at high urea concentrations at pH $>$ 5.5. Based on the observation that the
folding kinetics is mono-phasic, with no detectable burst phase in the initial fluorescence of the initial unfolding time course, it was surmised that there is no low energy intermediate in the unfolding of PDZ2. Rather the data were used to suggest the presence of a high-energy on-pathway intermediate,
which does not accumulate significantly in equilibrium \cite{Gianni05PEDS}. (iii) The high energy intermediate is non-detectable under stabilizing conditions,  achievable in PDZ2 domain by addition of modest amount of sodium sulfate. Under these conditions PDZ2 folds thermodynamically and kinetically in a two-state manner. (iv) The structures of the two transition state ensembles were also inferred using measured $\Phi$ values  as constraints in all atom molecular dynamics simulations \cite{Gianni07PNAS}. Unlike the results summarized in (i)- (iii) the predicted structures of the transition state ensembles are not as conclusive for reasons explained later in this work. 

In order to provide a comprehensive picture of folding of PDZ2, with potential implications for other single domain proteins,  we performed
molecular simulations of a coarse-grained off-lattice
model with side chains \cite{Hyeon11NatComm,Pincus08ProgMolBiolTranslSci} and used MTM
 \cite{O'Brien08PNAS,Liu11PNAS,Reddy12PNAS} to account for for denaturant effects.  The free energy profiles as a function of the structural overlap
order parameter at different urea concentrations,
$[C]$, and temperature $T$ suggest that the folding mechanism of
PDZ2  can be altered by changing the stability of the folded state.
In accord with experiments, we demonstrate directly the existence of the fleeting
 obligatory intermediate both in equilibrium ($I_{EQ}$) and  kinetics ($I_1$) \cite{Ivarsson07JBC}. The structures of $I_{EQ}$ and $I_1$ are similar. However, the
fraction of molecules in intermediate basin of
attraction (IBA),$f_{IBA}$,
 as a function of temperature and
$[C]$ is small, thus explaining  the difficulty in detecting it
 in standard denaturation experiments. In addition to  $I_1$
a kinetic intermediate, $I_2$, is consistently populated in $\sim$ 53\% of the folding trajectories. Guided by the free energy profiles, we identified two transition state ensembles, $TSE1$ and $TSE2$. The
computed values of the Tanford-like $\beta$ parameters, using the solvent accessible surface area as a surrogate, for the two
transition state ensembles (one connecting the NBA and the IBA and
the other involving transition between the IBA and the UBA) are in
qualitative agreement with those obtained from
experiments \cite{Gianni05PEDS,Ivarsson07JBC}.  The current work
further establishes that simulations based on the MTM are
efficacious in providing a nearly quantitative picture of folding of
single domain proteins.

\section*{\Large{METHODS}}

\textbf{SOP-sidechain model}: The simulations were carried out using a CG model in
which the $C_{\alpha}$-based self-organized polymer (SOP)
representation\cite{Hyeon06Structure} was augmented to include
side chains (SCs)\cite{Liu11PNAS,Liu12JPCB}. In the 
SOP-SC model each
residue is represented by two interaction centers, one that is
located at the C${_\alpha}$ position and the other  at the center
of mass of the side chain. In the SOP-SC model the native state
stabilization is achieved by accounting for backbone-backbone (bb),
side chain-side chain (ss), and backbone-side chain (bs) interactions present in the folded state. Neglect of non-native interactions, which do not significantly alter the folding mechanism beyond the global collapse of the protein \cite{Camacho95PROTEINS,Klimov01Proteins,Gin09JMB,Best13PNAS}, is nevertheless a limitation of the model.

The energy  (to be interpreted as an effective free energy obtained by
integrating over solvent (water) degrees of freedom) of a
conformation, describing the intra peptide interactions, is
\begin{mathletters}
\begin{equation}
E_P(\{r_i\})  =  V_{FENE} + V_{LJ}^{NAT} + V^{NEI} + V_{LJ}^{NN}.
\label{eq:E_P}
\end{equation}
The finite extensible nonlinear elastic potential (FENE),
$V_{FENE}$, accounting for  the chain connectivity between backbones
and side chains, is,
\begin{eqnarray}
V_{FENE} & = & V_{FENE}^{bb} + V_{FENE}^{bs} \nonumber \\
         & = & -\sum_{i=1}^{N-1}{k\over 2}R_{o}^2log(1-{(r_{i,i+1\_bb}-r_{i,i+1\_bb}^o)^2\over
         R_o^2}) \nonumber \\
         &   & -\sum_{i=1}^{N}{k\over 2}R_{o}^2log(1-{(r_{i,i\_bs}-r_{i,i\_bs}^o)^2\over
         R_o^2}).
         \label{FENE}
\end{eqnarray}

The non-bonded native interaction,  $V_{LJ}^{NAT}$ in Eq. (1) is taken to be
\begin{eqnarray}
V_{LJ}^{NAT} & = & V_{LJ\_NAT}^{bb} + V_{LJ\_NAT}^{ss} +
V_{LJ\_NAT}^{bs} \nonumber \\
             & = &\sum_{i=1}^{N-3}\sum_{j=i+3}^N e_{bb} \left[ \left({r_{i,j\_bb}^o\over r_{i,j\_bb}}\right)^{12}-2\left({r_{i,j\_bb}^o\over r_{i,j\_bb}}\right)^6\right]\Delta_{ij}^{bb}   \nonumber \\
             & + &\sum_{i=1}^{N-3}\sum_{j=i+3}^N e_{ss} |\epsilon_{ij}-0.7| \left[ \left({r_{i,j\_ss}^o\over r_{i,j\_ss}}\right)^{12}-2\left({r_{i,j\_ss}^o\over r_{i,j\_ss}}\right)^6\right]\Delta_{ij}^{ss}   \nonumber \\
             & + &\sum^N_{\begin{subarray}{l}i=1, j=1 \\ {|i-j|\geq 3}\end{subarray}} e_{bs} \left[ \left({r_{i,j\_bs}^o\over r_{i,j\_bs}}\right)^{12}-2\left({r_{i,j\_bs}^o\over
             r_{i,j\_bs}}\right)^6\right]\Delta_{ij}^{bs}.
\label{Non}
\end{eqnarray}
In Eqs. \ref{FENE} and \ref{Non} the superscript $^o$ refers to distances in the native state. If the distance between two non-covalently linked beads,
$r_{ij}$$(|i-j|\geq 3)$ in the PDB structure is within a cutoff
distance $R_c$, a native contact is formed, and correspondingly
$\Delta_{ij}=1$. If $r_{ij}$ exceeds $R_c$ then $\Delta_{ij}=0$. The strengths of the
non-bonded interactions $e_{bb}$, $e_{ss}$, $e_{bs}$ are assumed to
be uniform. The Betancourt$-$Thirumalai
(BT) \cite{Betancourt99ProtSci} statistical potential matrix
with elements $\epsilon_{ij}$, is used to explicitly treat the sequence
dependence.

We used repulsive interactions for excluded volume
effects between neighboring beads with strength $e_l$. The ranges of
repulsion are $\sigma_{bb}, \sigma_{i,j\_ss}, \sigma_{j\_bs}$ for
bb, ss and bs interactions respectively. The form of $V^{NEI}$ is
\begin{eqnarray}
V^{NEI} & = & V_{NEI}^{bb}+V_{NEI}^{ss}+V_{NEI}^{bs}\nonumber\\
            & = & \sum_{i=1}^{N-2}e_l\left( {\sigma_{bb}\over
            r_{i,i+2\_bb}}\right)^6 \nonumber\\
            & + & \sum_{i=1}^{N-1}e_l\left( {\sigma_{i,i+1\_ss}\over
            r_{i,i+1\_ss}}\right)^6 + \sum_{i=1}^{N-2}e_l\left( {\sigma_{i,i+2\_ss}\over
            r_{i,i+2\_ss}}\right)^6 \nonumber\\
            & + & \sum^N_{\begin{subarray}{l}i=1, j=1 \\ {0<|i-j|< 3}\end{subarray}}e_l\left( {\sigma_{j\_bs}\over
            r_{i,j\_bs}}\right)^6.
\end{eqnarray}
To prevent interchain crossing, we choose $\sigma_{bb}=a=3.8\AA$
($a$ is average distance between neighboring $C_{\alpha}$ atoms),
$\sigma_{i,j\_ss}=f(\sigma_i +\sigma_j)$ ($\sigma_i, \sigma_j$ are
the van der Waals radii of the side chains and $f=0.5$),
$\sigma_{j\_bs}=f(a +\sigma_j)$.

The non-bonded nonnative interactions are given by
\begin{eqnarray}
V_{LJ}^{NN} & = & V_{LJ\_NN}^{bb} + V_{LJ\_NN}^{ss} + V_{LJ\_NN}^{bs}\nonumber\\
            & = & \sum_{i=1}^{N-3}\sum_{j=i+3}^N
            e_l\left({ \sigma_{bb}\over r_{i,j\_bb}}\right)^{6}
            (1-\Delta_{ij}^{bb}) \nonumber\\
            & + & \sum_{i=1}^{N-3}\sum_{j=i+3}^N
            e_l\left({ \sigma_{i,j\_ss}\over r_{i,j\_ss}}\right)^{6}
            (1-\Delta_{ij}^{ss}) \nonumber\\
            & + & \sum^N_{\begin{subarray}{l}i=1, j=1 \\ {|i-j|\geq 3}\end{subarray}}
            e_l\left({ \sigma_{j\_bs}\over r_{i,j\_bs}}\right)^{6}
            (1-\Delta_{ij}^{bs}).
\end{eqnarray}
\end{mathletters}

Besides the knowledge-based BT statistical potential, the SOP-SC
energy function $E_P(\{r_i\})$ has seven parameters: $R_o=2\AA$,
$k=20kcal/(mol\cdot \AA^2)$, $R_c=8\AA$, $e_{bb}=0.73kcal/mol$,
$e_{bs}=0.17kcal/mol$, $e_{ss}=0.3kcal/mol$, $e_l=1kcal/mol$. Among
these parameters, $R_o$ and $k$ merely account for chain
connectivity.   The results would not be significantly affected by the precise choice of parameters enforcing the integrity of the polypeptide chain. Thus, in effect there are  five parameters in the SOP-SC model. Analysis
of PDB structures shows that $R_c=8\AA$ is a reasonable choice.  By analyzing structures of folded proteins in the Protein
Data Bank (PDB) that there are large favorable bb and bs
contacts \cite{Dima04JPCB}. The first and third terms account for
these interactions, which in turn ensures that packing effects are
appropriately described in the SOP-SC model. The experimental melting temperature of the protein is
 used to determine the  strengths of the native contacts, specified by
$e_{bb},e_{bs},e_{ss}$. 

\textit{\textbf{Molecular Transfer Model}}: The MTM theory \cite{Liu12JPCB} shows that
experimentally measured transfer free energies for
backbone and side chains along with the SOP-SC model could be utilized
to obtain the partition function for a protein at a finite denaturant
concentration [C]  \cite{Liu12JPCB}. In the MTM, the free energy of transferring a
given protein conformation, described by $\{r_i\}$ from water ([C]=0)
to aqueous denaturant solution ([C]$\neq$0), is approximated as,
\begin{equation}
\Delta G(\{r_i\},[C])  =  \sum_i \delta
g(i,[C])\alpha_i/\alpha_{Gly-i-Gly}
\end{equation}
where the sum is over backbone and side chain, $\delta g(i,[C])$ is
the experimentally measured transfer free energy of group $i$, $\alpha_i$ is the solvent
accessible surface area (SASA), and $\alpha_{Gly-i-Gly}$ is the SASA
of the $i^{th}$ group in the tripeptide $Gly-i-Gly$. Thus, the
effective free energy function for a protein at [C]$\neq$0 is
\begin{equation}
H_P(\{r_i\},[C]) = E_P(\{r_i\}) + \Delta G(\{r_i\},[C]).
\end{equation}
For computational expediency we
combined converged simulations at finite temperature using
$E_P(\{r_i\})$ and computed the partition function at [C]$\neq$0 to
obtain thermodynamic properties using a weighted histogram analysis
method, which takes into account the effects of $\Delta
G(\{r_i\},[C])$. Such an approximation, whose validity for obtaining
thermodynamic properties has been previously established \cite{Liu11PNAS}, is used to obtain
thermodynamic properties.

\textbf{ Langevin and Brownian Dynamics Simulations}:
We assume that the dynamics of the protein is governed by
the Langevin equation, which includes a damping term with a friction
coefficient $\zeta$, and a Gaussian random force $\Gamma$. The
equation of motion for a generalized coordinate $r_i$ is
$m\text{\"{r}}_i=-\zeta \text{\.{r}}_i +F_c+\Gamma$ where $m$ is the
mass of a bead, $F_c=-\partial E_P(\{r_i\})/\partial r_i$, is the
conformational force calculated using Eq. (\ref{eq:E_P}), $\Gamma$
is the random force with a white noise spectrum. The autocorrelation
function for $\Gamma(t)$ in the discretized form is
$<\Gamma(t)\Gamma(t+nh)>={2\zeta k_BT\over h}\delta_{0,n}$
\cite{Veitshans97F&D} where $\delta_{0,n}$ is the Kronecker
delta function and $n = 0, 1, 2 \ldots $. The value of the time
step, $h$, depends on the friction coefficient $\zeta$.

To obtain enhanced sampling, we used the  Replica Exchange Molecular
Dynamics
(REMD) \cite{Eisenmenger96CPL,Okamoto95JPC,Zhou01PNAS,Sugita99CPL} to
perform thermodynamics sampling using a low friction coefficient
$\zeta=0.05m/\tau_L$ \cite{Honeycutt92BP}, which allows us
to accurately calculate the equilibrium properties.  Rapid convergence of thermodynamics is possible
at a small $\zeta$ (under damped limit) because the polypeptide chain makes frequent
transitions between all accessible states. In the low $\zeta$ limit limit,
we used the Verlet leap-frog algorithm to integrate the equations of
motion. The velocity at time $t+h/2$ and the position at time $t+h$
of a bead are given by,
\begin{eqnarray}
v_i(t+h/2) & = & {2m-h\zeta\over 2m+h\zeta}\cdot v_i(t-h/2)+{2h\over 2m+h\zeta}\left[F_c(t)+\Gamma(t)\right] \\
r_i(t+h)   & = & r_i(t)+h\cdot v_i(t+h/2)
\end{eqnarray}

In order to simulate the kinetics of folding, $\zeta$ is set to be
$50m/\tau_L$, which approximately corresponds to the value in water
and represents the over damped limit \cite{Veitshans97F&D}. In
the high $\zeta$ value, we use the Brownian dynamics
algorithm \cite{Ermak78JCP}, which allows us to integrate
equations of motion using
\begin{equation}
r_i(t+h)=r_i(t)+{h\over \zeta}(F_c(t)+\Gamma(t)).
\end{equation}

\textbf{Time Scales}:
The natural unit of time for over damped condition at the transition
temperature $T_{s}$ is
$\tau_H \approx {\zeta_{H}a^2\over
k_BT_{s}}={(\zeta_H\tau_L/m)e_l\over k_BT_{s}}\tau_L$.
To convert the simulation time to real time, we chose $e_l=1kcal/mol$,
average mass $m=1.8 \times 10^{-22}g$ \cite{Veitshans97F&D},
$a=4\AA$, which makes $\tau_L=2ps$. For $\zeta_H= 50m/\tau_L$, we
obtain $\tau_H=159ps$. For thermal folding simulations, the
integration time step, $h$ is $0.005\tau_L$. In the
kinetic folding simulations, $h$, in Eq.(10) is $0.02\tau_H$.

\textit{\textbf{Data Analysis}}: The [C]-dependent melting temperature is identified with the peak in
the heat capacity.  The
structural overlap function $\chi=1-{N_k\over N_T}$ \cite{Camacho93PNAS} is used to
monitor the folding reaction, where
\begin{mathletters}
\begin{eqnarray}
N_k &=&\sum_{i=1}^{N-3}\sum_{j=i+3}^N
\Theta\left(\delta-|r_{i,j\_bb}-r_{i,j\_bb}^o|\right)
+\sum_{i=1}^{N-3}\sum_{j=i+3}^N\Theta\left(\delta-|r_{i,j\_ss}-r_{i,j\_ss}^o|\right)\nonumber\\
&+&\sum^N_{\begin{subarray}{l}i=1, j=1 \\
{|i-j|\geq 3}\end{subarray}}
\Theta\left(\delta-|r_{i,j\_bs}-r_{i,j\_bs}^o|\right)
\end{eqnarray}
\end{mathletters}
In Eq.(11), $\Theta(x)$ is the Heavyside function. If
$|r_{i,j}-r_{i,j}^o| \leq \delta$ (=$2\AA$), there is a contact. The
number of contacts in the $k^{th}$ conformation is $N_k$, and $N_T$
is the total number in the folded state.

\section*{\Large{Results}}
\textbf{Thermal Denaturation:}:  The temperature dependence of the heat capacity, $C_v(={<E^2>-<E>^2\over k_BT^2})$,
where $<E>$ and  $<E^2>$ are the mean and mean square averages of the energy,
respectively, demonstrates that PDZ2 folds cooperatively in a
two-state manner (Figure 1b). The melting temperature, identified with
the peak in $C_v(T)$,  is $T_m=324K$. The value of $T_m$ obtained in
our simulations is in excellent agreement with the experimentally
measured $T_m=321K$ \cite{Ivarsson08PEDS}.  

In order to identify the $NBA$, $UBA$, and the intermediate basin of attraction ($IBA$) representing the $I_{EQ}$ state, we  plot in Figure 1c the free energy ($G(\chi)$)
profile as function of $\chi$ at  $T_m=324K$. All the conformations can be
classified into three states specified by the black vertical lines based on the $\chi$ values.
 If $\chi \leq \chi_c^N = 0.64$, the 
conformations are in the NBA, conformations with  $\chi \geq
\chi_c^D =0.825$ belong to the UBA, and  the rest of the
conformations are in the  IBA. The fractions of
molecules in the $NBA$, $f_{NBA}([0],T)$ (the
first argument shows the value of the denaturant concentration), in
the $UBA$ $f_{UBA}([0],T)$, and in the $IBA$, $f_{IBA}([0],T)$ as a function of
temperature are plotted in Figure 1d. Both $f_{NBA}([0],T)$ and
$f_{UBA}([0],T)$ show that the folding or unfolding is
cooperative. The value of $f_{IBA}([0],T)$ is negligible compared to
$f_{NBA}([0],T)$ and $f_{UBA}([0],T)$, suggesting that from a
thermodynamic perspective a  two-state description is adequate,
reflecting the cooperative transition in the heat capacity curve
(Figure 1b). The sparse population of the $IBA$  explains why the $I_{EQ}$
 state is hard to detect in experiments although its
presence appears as a shoulder in the free energy profile (Figure
1c). The value of $T_m$ computed using $f_{NBA}(T_m) = 0.5$ yields
$T_m = 324K$, which coincides with the peak in the heat capacity
(Figure 1b).

\textbf{Chemical Denaturation:}  Following our
previous studies \cite{O'Brien08PNAS,Liu11PNAS,Reddy12PNAS}, we
choose a simulation temperature, $T_s$, at which the calculated free
energy difference between the  native state (N) and the
unfolded state (U), $\Delta G_{NU}(T_s)$ ($G_N(T_s)-G_U(T_s)$) and
the measured free energy $\Delta G_{NU}(T_E)$ at $T_E$ (=298K)
coincide. The use of $\Delta G_{NU}(T_s)=\Delta G_{NU}(T_E)$ ( in
water with $[C]=0$ ) to fix $T_s$ is equivalent to choosing a
overall reference energy scale in the simulations. For PDZ2, $\Delta
G_{NU}(T_E=298K)=-3.1 kcal/mol$ at $[C]=0$ \cite{Sicorello09BJ},
which results in $T_s=317K$. Besides the choice of $T_s$, {\it no other
parameter} is adjusted to obtain agreement with experiments for any property.

With $T_s=317K$ fixed, we calculated the
dependence of $f_{NBA}([C],T_s)$, $f_{UBA}([C],T_s)$, and
$f_{IBA}([C],T_s)$ on $[C]$ (Figure 2a). The agreement between the
measured and simulated results for $f_{NBA}([C],T_s)$ as a function
of  [C] is excellent (Figure 2a).  We find,
just as in thermal denaturation, that $f_{IBA}([C],T_s)$ is small
\cite{Gianni05PEDS,Sicorello09BJ}. The midpoint concentration,
$C_m$, obtained using $f_{NBA}([C_m],T_s)=0.5$ is [C]=2.3M, agrees
well with the experimentally measured value of 2.6M (see Figure 6D in \cite{Sicorello09BJ}).

The native state stability with respect to U, $\Delta
G_{NU}([C])$($=G_N([C])-G_U([C])$), is computed using $\Delta
G_{NU}([C])=-k_BT_sln({f_{NBA}\over f_{UBA}})$. The linear fit,
$\Delta G_{NU}([C])=\Delta G_{NU}(0)+m[C]$, yields $\Delta
G_{NU}([0])=-3.09kcal/mol$ and $m=1.35kcal/mol\cdot M$ (Figure 2b).
The experimentally inferred $m = 1.20kcal/mol\cdot M$
compares well with the simulations, which establishes again
that simulations based on  MTM predict the  thermodynamic
properties  of proteins accurately.
The $[C]$-dependent heat capacity curves (Figure 2c)
show that the peaks corresponding to $T_m([C])$ decreases as $[C]$
increases (Figure 2c=d). Taken together the simulations
show that the  equilibrium folding induced by temperature or
denaturants is cooperative.

{\bf Urea-dependent changes in the shape of PDZ2:} The dependence of the radius of gyration, $\langle R_g([C]) \rangle$, on urea concentration (black line in Figure 3a) shows a transition from an expanded to a collapsed state as $[C]$ decreases. The radii of gyration of the $I_{EQ}$ and the native state are virtually constant at all urea concentrations. However, the radius of gyration of the $UBA$ decreases as $[C]$ decreases implying that the ensemble of structures of the unfolded state is more compact under native conditions than at 8M urea. The decrease in the $R_g$ of the $UBA$ structures in going from 8M urea to 1M is about 9\%, which should be measurable in a high precision Small Angle X-ray scattering (SAXS) experiment (see below for additional discussion). The $P(R_g)$ distributions at various urea concentrations show the expected behavior (Figure 3b). The protein is largely in the $NBA$ at $[C]$ less than 2.3M (mid point of the folding transition)  and is expanded at higher values of $[C]$. The distance distributions, which can be measured as the inverse Fourier transform of the wave vector dependent scattering intensity, are plotted in Figure 3c and constitutes one of the predictions.

{\bf Free energy profiles, $G(\chi)$, reveal lowly populated intermediate:} To illustrate how  urea changes the folding
landscape, we plotted $G(\chi)$ versus $\chi$ at different $[C]$ at $T=T_m=324K$ in Figure 4a and at
$T=T_s=317K$ in Figure 4b.  Figure 4a shows, that at all urea
concentrations,  the conformations could be partitioned into
three states, which is consistent with the results in Figure 1c displaying $G(\chi)$ at $T_m$ with $[C] = 0$. The basin of attraction
corresponding to $I_{EQ}$ becomes deeper as $[C]$
increases but is  shallow compared to the $NBA$ and $UBA$.  The barrier for $TSE1$ is  lower than for $TSE2$, implying that the transition from the intermediate states to the
UBA is more facile than transition to the $NBA$. The number of distinct minima
remain unchanged at the lower temperature, $T_s=317K$
(Figure 4b). However, the basin for the intermediate states becomes
 shallower as $[C]$ increases, and the barrier for TSE1
almost disappears, especially when $[C]>C_m=2.3M$, which indicates
that the intermediate, if it can be detected at all, is unstable. The absence of $I_{EQ}$ at the lower temperature suggests that stabilization of the native fold of PDZ2 leads to a two-state thermodynamic transition. Our finding supports the same observation in experiments showing that native state stabilization upon addition of sodium sulfate results in the high energy $I_{EQ}$ being undetectable. 

\textbf{Structures of the transition state ensembles ($TSE1$ and $TSE2$):} We identified the
transition state structures using the folding trajectories generated  at $T_m$. We picked the  putative transition state from the barrier regions of the free energy
profile as a function of $\chi$, using the conditions $ 0.815<\chi <
0.835$ for $TSE1$ and $ 0.63<\chi < 0.65$ for $TSE2$, which are
represented as shaded areas in Figure 1c.  Starting from these
structures, we calculated the commitment probability, $P_{fold}$, of
reaching the $NBA$ \cite{Du98JCP}. The sets of structures with $ 0.4 \leqslant
P_{fold} \leqslant 0.6$ are identified as the transition state
ensemble. The characteristics of the two transition state ensembles are displayed in Figure 5a and the distribution of the structural order parameter $P(\chi)$ of the $TSE1$ along with the contact map are shown in Figure 5b and Figure 5c, respectively. Distribution of $P(\chi)$ and the structural details for $TSE2$ are shown in Figures 5d and 5e, respectively.

The  characteristics of the TSEs are experimentally described
using the Tanford $\beta$ parameter. By fitting the  measured Chevron plots as linear function of $[C]$
it has been shown that $\beta_E^1=0.53$ for TSE1, and
$\beta_E^2=0.89$ for TSE2 \cite{Gianni05PEDS} and $\beta_E^1=0.35$
for TSE1 and $\beta_E^2=0.85$ for TSE2 \cite{Ivarsson07JBC}.  It is
generally assumed that $\beta$ is related to the buried solvent
accessible surface area (SASA) in the transition state. For the
$TSE$s obtained in our simulations, we calculated the distribution
$P(\Delta_R)$ (see Figure 5a), where
$\Delta_R=(\Delta_U-\Delta_{TSE})/(\Delta_U-\Delta_N)$ with
$\Delta_U$, $\Delta_{TSE}$, $\Delta_N$ being the SASAs  in the DSE
($[C]=8.0M$), TSE, and the NBA ($[C]=0.0M$), respectively. We found
that the average $<\Delta_R>=0.23=\beta_s^1$ for $TSE1$ and
$<\Delta_R>=0.68=\beta_s^2$ for $TSE2$, which qualitatively agree
 with the values inferred from experiments. The small value of $\beta_1$ suggests that the $TSE1$ structures are more $UBA$-like whereas the $TSE2$ structures resemble the native state.

The structural details are revealed in Figure 5c, which shows the contact map for $TSE1$. It is clear
that relative to the native state (top half in Figure 5c) the extent to which
the structure is ordered in $TSE1$ is modest (green) to
low (blue). The contacts between $\beta$ strands $1-6-4$ have
low formation probabilities as indicated by the two black circled
regions.
The secondary structures, $\beta_{23}$ and two helices, are ordered
to a greater extent. A representative structure for $TSE1$ displayed  on
the left of Figure 5a  shows that the structure is expanded.
Only $\beta_{23}$ and the two helices have relatively
high formation probability. 

The contact map for $TSE2$ (Figure 5e) shows that the formation 
probabilities of contacts even between residues that are distant in sequence are high, which results in the ensemble of   TSE2 structures being 
compact. The major blue region in the contact map
indicates that $\beta_{16}$ is still largely unstructured. Four
superimposed representative structures from $TSE2$ are shown on the
right of Figure 5a. The structures are native-like
except that $\beta$ strand 1 is not as well packed as in the native state. The lack of stabilizing interactions in 
$\beta_{16}$ found in our simulations  disagrees with the inferences from the all atom 
molecular dynamics simulations using measured $\Phi$ values as constraints \cite{Gianni07PNAS} (see below for additional discussion).

\textbf{Folding and collapse kinetics:} We calculated the collapse and folding rates at zero urea concentration from the 
folding trajectories , which were generated from Brownian
dynamics simulations \cite{Veitshans97F&D}. From 93 folding trajectories, the fraction
of unfolded molecules at time $t$, is computed using
$P_u(t)=1-\int_0^tP_{fp}(s)ds$, where $P_{fp}(s)$ is the
distribution of first passage times. We fit $P_u(t) = e^{-tk_f}$ (
Figure 6a) with $k_f=172s^{-1}$, a value that is
about 60 times larger than found in the experiments ($2.8^{-1}$) \cite{Gianni05PEDS}. The kinetics of collapse of PDZ2
domain, shows that $<R_g(t)>$ decays with a single rate
constant, $k_c([C])$, the rate of collapse. The extracted values of
$k_c([C])=244s^{-1}$ from the data in Figure 6b is greater than
$k_f([C])$, which shows compaction occurs ahead of folding. 

{\bf Thermodynamic and kinetic intermediates:} Analyses
of the folding trajectories show that the folded state is reached through a
kinetic intermediate state, $I_1$ (Figures 7a, 7b, 7c, and 7d). This intermediate state, while not  directly detected
in experiment, is likely responsible for the downward curvature observed in  the unfolding arm of the  chevon
plots \cite{Gianni05PEDS}. We also find the presence of another kinetic intermediate
states ($I_2$)
 in some  of the trajectories. A quantitative analyses allows us to classify the occurrence of $I_1$ and $I_2$ in the $93$ folding trajectories. 
In $49$ trajectories both $I_1$ and $I_2$ are found.  Two
trajectories in which this occurs  are illustrated in Figures 7a and 7b.
In Figure 7a, PDZ2 samples  $I_1$ and $I_2$ only once before
reaching the native state whereas in Figure 7b, it samples $I_1$ and $I_2$
more than once. In the second class  of trajectories ($44$ out of
$93$) PDZ2 samples only the $I_1$ (see  Figures 7c and 7d for examples). Structures of $I_1$ and $I_2$ are displayed in Figure 7e.

{\bf $I_{EQ}$ and $I_1$ are structurally similar:}
To illustrate the structural similarity between $I_{EQ}$, identified in the free energy profiles (Figures 1 and 4), and $I_1$  in 100\% of the kinetic folding trajectories, and $I_2$ sampled in some of the folding trajectories, we computed the  average fraction of native contacts formed by
every residue, $f_{Q}$, for the three states (Figures 8a and 8b).  The correlation between $I_{EQ}$ and $I_1$, shown in Figure 8c, is very
high (R=0.995).   Therefore, we surmise that $I_1$ and $I_{EQ}$  are structurally identical. 
 Quantitative results in Figures 8b and 8d and sample structures (Figure 8e) show that the  structure of $I_2$ differs from $I_{EQ}$. Taken together these results show that both thermodynamic and kinetic intermediates can be sampled during the folding process although the former is far more prevalent. 

\section*{\Large{Discussion:}}

{\bf Minimum Energy Compact Structures (MECS):} The folding trajectories reveal that two major intermediates are sampled as PDZ2 folds.  The equilibrium intermediate is found in all the trajectories whereas the kinetic intermediate is found in only a fraction of the trajectories. Two aspects of these findings, which are of general validity for folding, are worth pointing out. (i) Both $I_1$ and $I_2$ form on the time scale of collapse. In those trajectories in which $I_2$ forms there frequent transitions between $I_2$ and $I_1$ (see Figure 7). In the process of making such transitions PDZ2 undergoes considerable expansion.  (ii) The intermediates, $I_{EQ}$ and $I_1$, are compact and contain  native-like features. The major difference between $I_1$ and the folded state is in the extent of structure in $\beta_1$, $\beta_4$, and $\beta_6$ as well as $\alpha_2$. The structure of $I_1$ is similar to that found in $TSE1$, which follows from the Hammond postulate.    These intermediates, which are like MECS (minimum energy compact structures) \cite{Camacho93PRL}  facilitate rapid folding. Although they are difficult to detect in ensemble experiments, single molecule pulling experiments using cycles of force increase and force quench can be used to detect MECS as has been done for Ubiquitin \cite{Garcia-Manyes09PNAS}. It would be interesting to do similar experiments on PDZ2 to directly observe $I_1$ and $I_2$ using force as a perturbation.

{\bf Collapse and folding transition:}  Is the size of the polypeptide chain of the unfolded state under folding conditions ($[C] < C_m$)  less than at $[C] > C_m$?  Theoretical arguments \cite{Camacho93PNAS,Ziv09JACS,Ziv09PCCP}, simulations \cite{O'Brien08PNAS}, and a number of single molecule FRET and SAXS experiments \cite{Schuler08COSB,Haran12COSB} have answered the question in the affirmative whereas some SAXS experiments on  protein L  suggest that there is no evidence for  polypeptide chain  collapse, which is manifestly unphysical. The arguments in favor of collapse preceding folding is based on the following observations. The random coil state of a polypeptide chain with $N$ residues is expected to be $R_g^{U} \sim a_R N^{0.6}$ with $a_R \approx$ 2.0\AA. This estimate  is likely to be an upper bound because even 8M urea is not a good solvent because even at these elevated concentrations the unfolded state has residual structure. As the denaturant concentration decreases the maximum decrease in $R_g$ is likely to have a lower bound $a_D N^{0.5}$. It cannot be maximally compact because if it were so then enthalpic interactions would drive these structures to the folded state for which $R_g \approx a_N N^{1/3}$ with $a_N$ =3.0\AA \cite{dima04JPCB}. Thus, we surmise that $  a_D N^{0.5} < R_g^{U} <   a_R N^{0.6}$. The reduction in $R_g$ of the unfolded state for $N= 64$ (protein L) is predicted to be between (5-10)\% depending on the values of $a_R$ and $a_D$. In PDZ2 ($N = 94$) we find that the unfolded state $R_g$ decreases by only 9\% as $[C]$ decreases (Figure 3a). Thus, due to finite $N$ resulting in a small decrease in the unfolded state $R_g$ requires high precision SAXS measurements to measure changes in $R_g^U$ as $[C]$ decreases. The errors in SAXS for protein L are far too large to accurately estimate $R_g^U$, especially under native conditions. 

The absence of detectable contracted form of the polypeptide chain in time resolved SAXS experiments protein L was used as further evidence that compact states are not formed in the folding of any single domain protein. Using theory \cite{deGennes85JPL,Pitard98EPL,Thirum95JPI} it can be shown that collapse of the unfolded state occurs on time scales on the order of $\tau_c \approx \tau_0 N^{\beta}$ where $\beta$ is between 1.5 and 2.2, and $\tau_0 \approx \mathcal O {10^{-9} s}$. For PDZ2  we estimate that $\tau_c \sim 20 \mu s$. Our simulations show that the contraction of the unfolded state occurs on time scale that does not exceed a maximum of $\approx 50 \mu s$ (see the inset in Figure 6b). Theoretical estimate based on the scaling law above for collapse of the unfolded state of protein L ($N = 64$) a maximum of $\tau_c \sim 4 \mu s$. Based on reconfiguration time measurements using Fluorescence Correlation Spectroscopy \cite{Nettels07PNAS,Soranno12PNAS} indicate that $\tau_c$ for small proteins could be less by an order of magnitude compared to the theoretical estimate.  A larger value of $20 \mu s$, much shorter than the folding time,  for reconfiguration time has also been reported for protein L \cite{Waldauer10PNAS}. For the larger DHFR ($N = 154$) there is nearly a 23\% reduction in $R_g^U$ in $\sim 300 \mu s$ \cite{Arai07JMB}, which is also considerably shorter than the folding time. All the studies show that collapse of the unfolded state, which increases with $N$, is on the order of at most tens of $\mu s$. Thus, we conclude that the time resolution in the most recent SAXS experiments on protein L (4 ms) \cite{Yoo12JMB}, which is comparable to the folding time, is too long to shed light on compaction of the polypeptide chain. The presence of 15 His tags in the first study \cite{Plaxco99NSB} makes it difficult to ascertain the relevance for protein L. Indeed, the inability to accurately determine determine the characteristics of the unfolded state had also lead to erroneous conclusions about equilibrium collapse and kinetic foldability \cite{Millett02Biochem}, which has been corrected recently using smFRET experiments recently \cite{Hofmann12PNAS}.  

Fast mixing experiments that simultaneously detect compaction and acquisition of structure on a number of proteins \cite{Chan97PNAS, Akiyama02PNAS,Kimura05PNAS} have produced ample evidence that collapse is an integral process of the folding process, as predicted by theory. Although it is likely that the simplified analysis of smFRET measurements overestimates the extent of collapse \cite{O'Brien09JCP}, fast mixing SAXS experiments  on a variety of proteins leave no doubt that the unfolded state is indeed more compact (albeit by only about $\approx$ 9\%) under native conditions despite persistent claims to the contrary based on SAXS data based largely on one protein (protein L) with large errors. It is a pity that the erroneous conclusion has resulted in unnecessary obfuscation. What is needed are high precision data for single domain proteins spanning a range of $N$ (say between 50-250), which cannot be easily obtained by  SAXS alone  \cite{Trewhella13Structure} but is more readily available in smFRET experiments.

{\bf Detection of $I_{EQ}$ in PDZ2 in single molecule pulling experiments:} The downward curvature in the unfolding rate as a function $[C]$ in PDZ2 implies that an intermediate is populated \cite{Gianni05PEDS}. Single molecule pulling experiments are well suited to explore this finding more readily.   Based on the free energy profile computed here and postulated elsewhere \cite{Ivarsson07JBC}, we suggest that unfolding by mechanical force ($f$) would give rise to downward curvature in a plot of $log k_u(f)$ versus $f$. At low forces the inner barrier separating the $NBA$ and $I_{EQ}$ would dominate whereas at high forces the second outer barrier is relevant for mechanical unfolding. The two sequential barrier picture implies that there would be two transition state distances with a switch  between the two occurring as $f$ is increased. A plausible support for this argument comes from the observation that the Tanford $\beta_T$ as a function of $[C]$ exhibits a sigmoidal behavior (see Figure 4 in \cite {Ivarsson07JBC}) with $\beta_T$ changing from 0.35 at low $[C]$ to 0.85 at high $[C]$. The scenario postulated here is distinct from that in SH3 domain in $[log k_u(f), f]$ plot exhibits an upward curvature, which is a signature of parallel unfolding pathways with a switch between the transition state ensembles \cite{Guinn15NatComm} as opposed to the predicted sequential barrier model for PDZ2. Laser optical tweezer experiments are ideally suited to test our prediction.

{\bf Structures of the TSEs:} It is interesting to compare the structures of TSE1 and TSE2 obtained in this work with those reported earlier \cite{Gianni07PNAS}. The structure of TSE1 (Figure 5) show interactions involving $\beta_2$, $\beta_3$, and formation of the two helices. This is  in sharp contrast to
the conclusion in  \cite{Gianni07PNAS} suggesting that in TSE1 $\beta$
strands $1-6-4$ are structured  whereas the other
secondary structural elements  are essentially disordered.  Although the agreement between the predicted structures in our work and those reported in \cite{Gianni07PNAS} in TSE2 is better in that we find that it is native-like, there are crucial differences as well. In particular, the lack of stabilizing interactions involving 
$\beta_1$ and $\beta_6$ found in our simulations  disagrees with the inferences drawn from the all atom 
molecular dynamics simulations using measured $\Phi$ values as constraints \cite{Gianni07PNAS}.  The differences are likely to be related to the completely different approaches used in the two studies. In \cite{Gianni07PNAS} the measured $\Phi$ values were used as constraints in standard all atom molecular dynamics simulations, which use inaccurate force fields. The procedure, while interesting, is not systematic in the sense the accumulation of errors both from the $\Phi$ values as well as the MD simulations is nearly impossible to quantify. More importantly, the putative TSE structures were not used to obtain the $P_{fold}$ values in the earlier study \cite{Gianni07PNAS}, which casts doubt on whether the identified TSEs accurately represent the actual TSEs. We believe additional experiments, including perhaps double mutant cycles, would be needed to ascertain the nature of the TSEs in PDZ2.    

\section*{\Large{Concluding Remarks:}}

We have used a phenomenological theory based on the MTM to simulate the folding of PDZ2 domain as function of temperature and urea. In addition to providing support to the folding mechanism discovered in experiments \cite{Gianni05PEDS}  we have made a few testable predictions. (1) We have obtained a precise dependence of the melting point as a function of urea concentration, which can tested using standard calorimetry experiments. (2) The presence of high energy intermediate in the absence of added salt can be characterized using single molecule pulling experiments as shown using simulations for srcSH3 domain \cite{Zhuravlev14JMB} demonstrating that  the excited state is sparsely ($O \approx {2-5)\%}$ populated, which coincided with the findings based on NMR experiments \cite{Neudecker12Science}.

\section*{\Large{Acknowledgements}}
We are grateful to Ben Schuler and Gilad Haran for their interest and useful comments. This work was supported in part by a grant from the National Science
Foundation to DT through CHE 13-61946. Z.L. acknowledges partial financial
support from the National Natural Science Foundation of China (11104015) and the Fundamental Research Funds for the Central
Universities (2012LYB08).


\newpage


\newpage
\section*{\large Figure Captions}
\hspace{-1.2em}Figure 1: Thermodynamics of folding. (a) Ribbon diagram
representation of PDZ2(PDB code: 1GM1). (b) Temperature dependence
of specific heat (blue) and total energy (magenta). (c) Free energy
profile at $T_m$ as a function of $\chi$. The values $\chi_c^N$ and
$\chi_c^D$ are used to classify the major equilibrium states. The shaded areas give putative regions for the two transition state ensembles. (d) Fraction of molecules in NBA(black), UBA(red) and IBA(green) as
functions of temperature.
\\
\\
\hspace{-1.2em}Figure 2: Denaturation effects. (a) Fraction of molecules in the
NBA (black), UBA (red), and IBA (green) as a function of urea
concentration $[C]$.  For comparison, the experimental curve for $f_{NBA}[C]$ (blue) is  shown. (b) $[C]$ dependence of
free energy of stability of the native state with respect to the
unfolded state. Fit to a linear function yields $\Delta G_{NU} = \Delta G_{NU}(0) + m[C]$ where $\Delta G_{NU}(0) = -3.09 kcal/mol$ and $m = 1.35 kcal/mol.M$.  (c) Heat capacity versus temperature
for different values of $[C]$. (d) The $[C]$ dependence of the melting temperature. The line is a fit to $T_m[C] = T_m(0) - B[C]$ where $T_m(0)$ = 324K and $B = -3.1 KM^{-1}$.
\\
\\

\hspace{-1.2em}Figure 3: Equilibrium collapse. (a) Average $<R_g>$ (black) as
a function of $[C]$. Red, green and blue curves correspond to $<R_g>$ of the
folded, unfolded and $I_{EQ}$ states, respectively. The scale for the unfolded state is on the right.
(b)Distribution $P(R_g)$ of $R_g$ for various
concentrations of urea. The inset shows $P(R_g)$ for $[C]=0$(black),
$[C]=2.3$(blue), and $[C]=5.0$(orange) corresponding to the extended
conformations ($R_g
> 16\AA$). (c) Distance distribution function $P(r)$, the inverse Fourier
transform of the scattering intensity, for 0M (black), 1.0M (red),
2.3M (green), and 5.0M (blue) urea. Here, $r$ is the distance between
all non-covalently linked beads.
\\
\\

\hspace{-1.2em}Figure 4: Free energy profiles versus $\chi$ at different
$[C]$. (a) $T=T_m$. (b) $T=T_s$. The values of $[C]$ measured in M
from top to bottom are 0, 1, 2, 2.3, 3, 4, 5, 6, 7, and 8.
\\
\\
\hspace{-1.2em}Figure 5: Quantifying the transition state ensembles. (a) Distribution
$P(\Delta_R)$, of the
$\Delta_R=(\Delta_U-\Delta_{TSE})/(\Delta_U-\Delta_N)$, which is the
fraction of buried solvent accessible surface area relative to the
unfolded structures. The average $<\Delta_R>=0.23$ for TSE1 and
$<\Delta_R>=0.68$ for TSE2. These values coincide qualitatively with
Tanford $\beta$ parameters extracted from the observed Chevron plot. A few of the TSE1 (TSE2) structures are displayed on the left (right) respectively.
(b)The distributions of $\chi$ computed from  400 simulation
trajectories spawned from the transition state structures in TSE1.
Data are shown for the four different structures. The distribution
shows that roughly half of these trajectories go to the folded basin
of attraction($P_{fold}\thickapprox 0.5$).(c) Contact map of the
native state ensemble (upper left) and the one for the TSE1(lower
right). The scale on the right gives the probability of contact
formation. (d) and (e) Same as (b) and (c), respectively  except the results  are for TSE2. 
\\
\\
\hspace{-1.2em} Figure 6: Folding and collapse kinetics.  
(a) Fraction of molecules that have not folded ($[C]=0.0M$) as a
function of time. The line is an exponential fit ($P_u(t) = e^-{t/\tau_F}$ with $\tau_F = 5.8 ms$) to the data.
(b) Collapse kinetics monitored by the time-dependent average
$<R_g(t)>$ as a function of $t$ with the line giving an exponential fit to the data. The collapse time of PDZ2 is $\tau_c = 4.1 ms$.
\\
\\

\hspace{-1.2em} Figure 7: Two major folding pathways monitored by $\chi$ as a function of $t$. (a) and (b) show two
representative trajectories showing that the native state is reached by sampling  both $I_1$ and $I_2$.  (c) and (d)
show two folding  trajectories in  which the polypeptide chain samples only $I_1$, often multiple times, before reaching the folded state. (e) Structures for I1 and I2.
\\
\\
\hspace{-1.2em}Figure 8: Comparing $I_{EQ}$, $I_1$, and $I_2$.  (a)Average fraction of native contacts formed for
residues, $f_Q$, for $I_{EQ}$ (black) and $I_1$ (red). (b) $f_Q$ for $I_{EQ}$ (black) and $I_2$ (blue). (c) Correlation between $f_Q$s for $I_{EQ}$ and $I_1$. The correlation coefficient is near unity. (d) Relation
between $f_Q$s for $I_{EQ}$ and $I_2$ with correlation coefficient $\approx 0.9$. (e) Structures of the three intermediates. 
\\

\hspace{-1.2em}Figure 9: Table of contents.

\newpage
\begin{figure}[ht]
\includegraphics[width=\columnwidth]{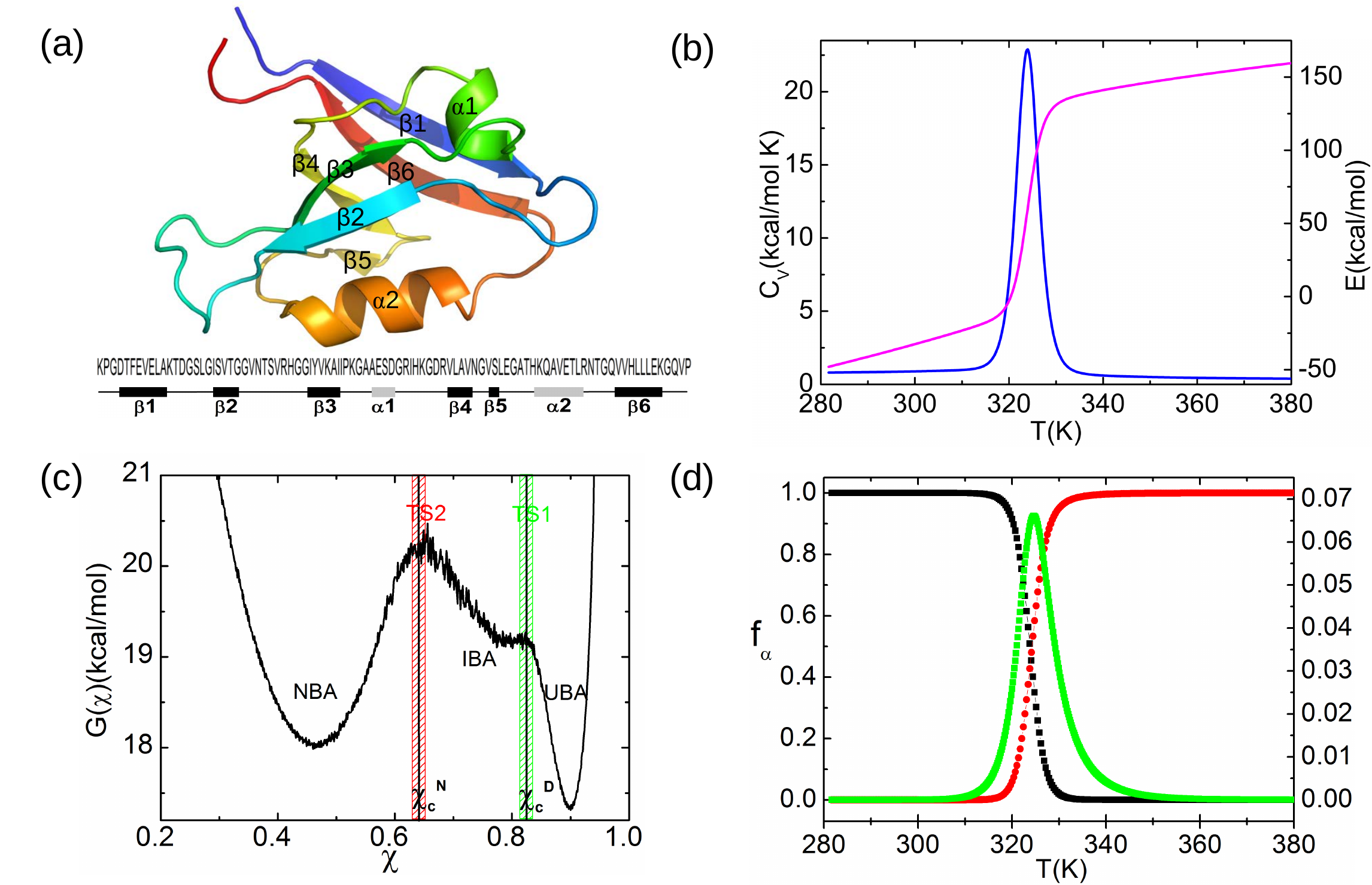}
\caption{}
\end{figure}
\clearpage

\begin{figure}[ht]
\includegraphics[width=\columnwidth]{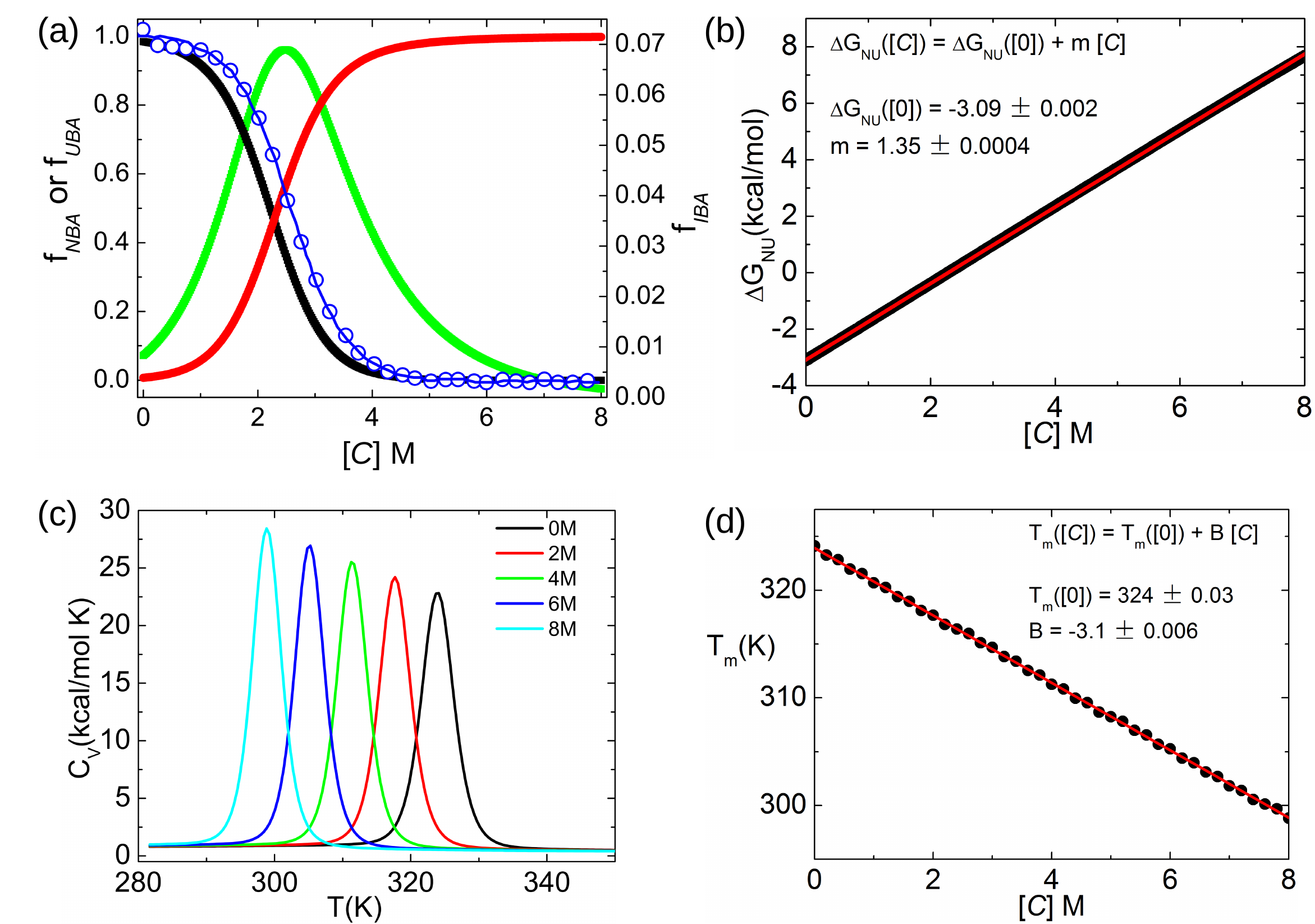}
\caption{}
\end{figure}
\clearpage

\begin{figure}[ht]
\includegraphics[width=0.5\columnwidth]{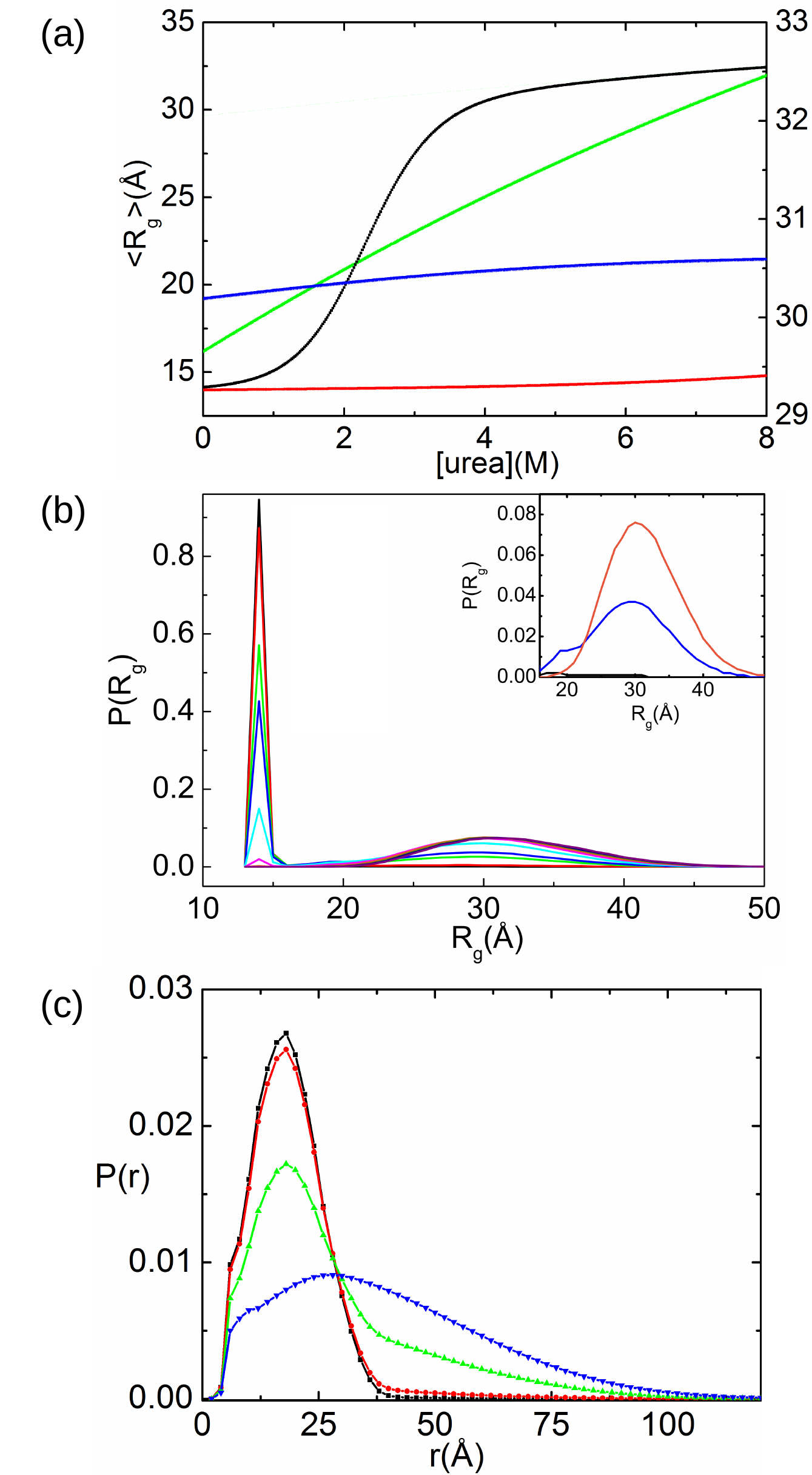}
\caption{}
\end{figure}
\clearpage

\begin{figure}[ht]
\includegraphics[width=\columnwidth]{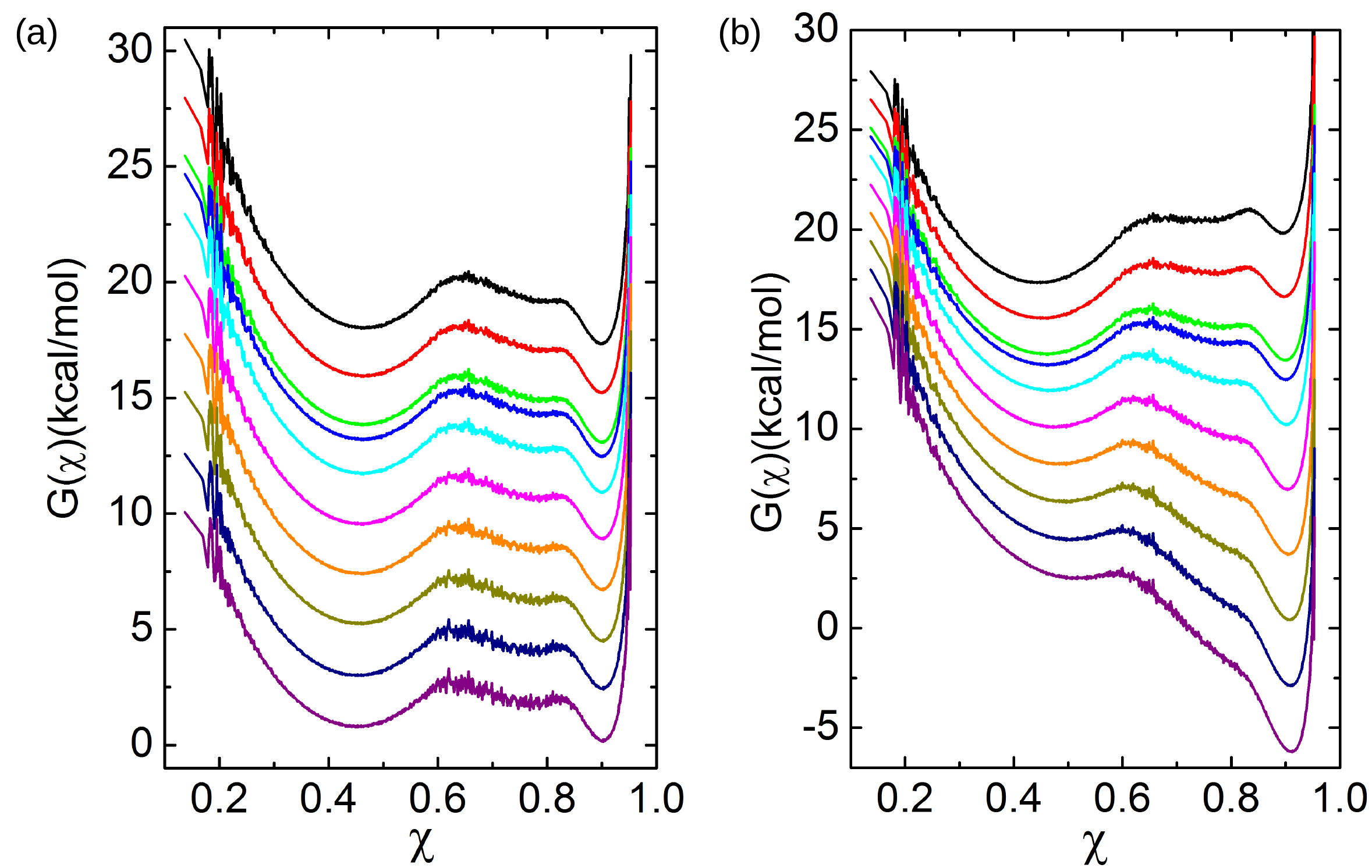}
\caption{}
\end{figure}
\clearpage

\begin{figure}[ht]
\includegraphics[width=\columnwidth]{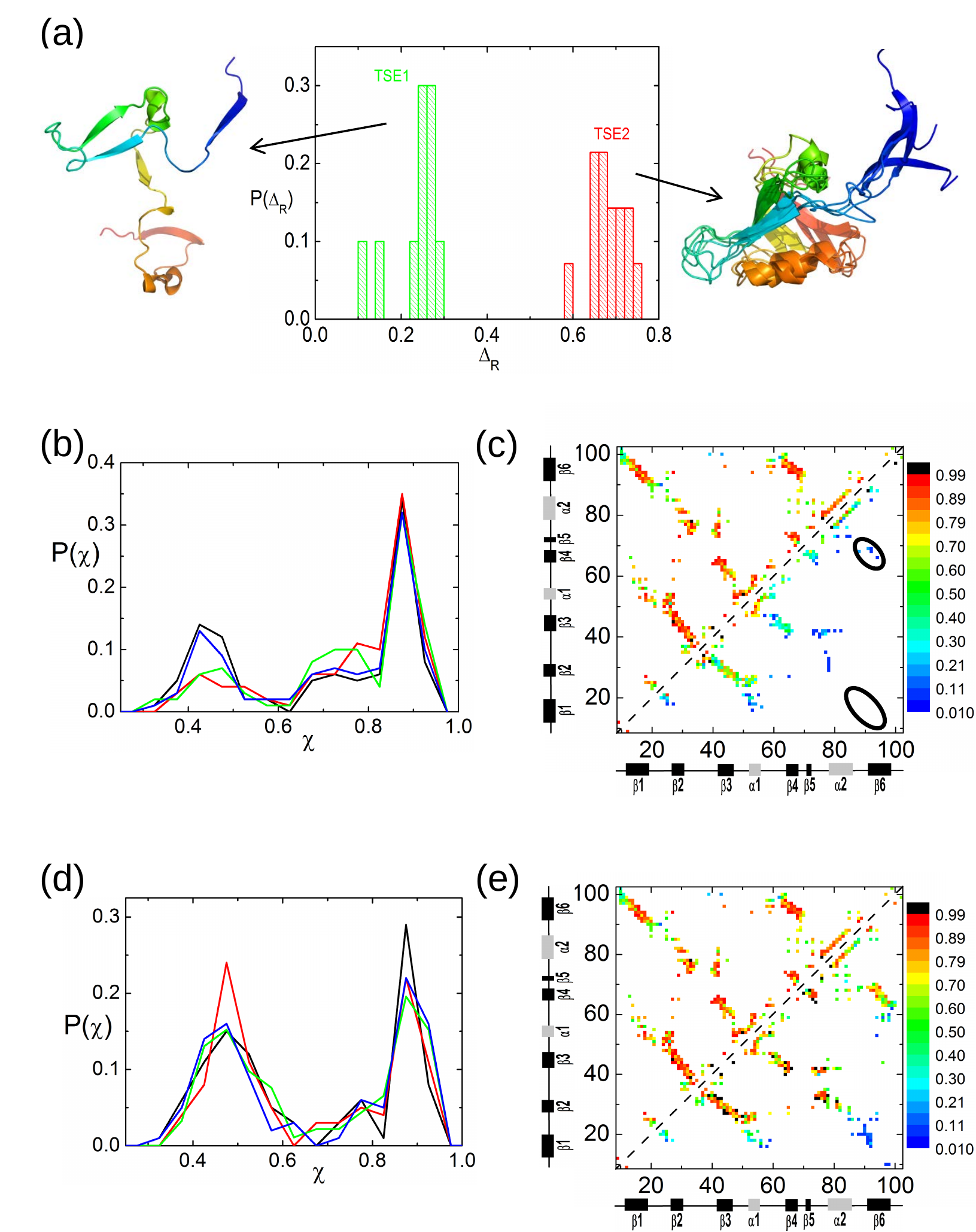}
\caption{}
\end{figure}
\clearpage

\begin{figure}[ht]
\includegraphics[width=\columnwidth]{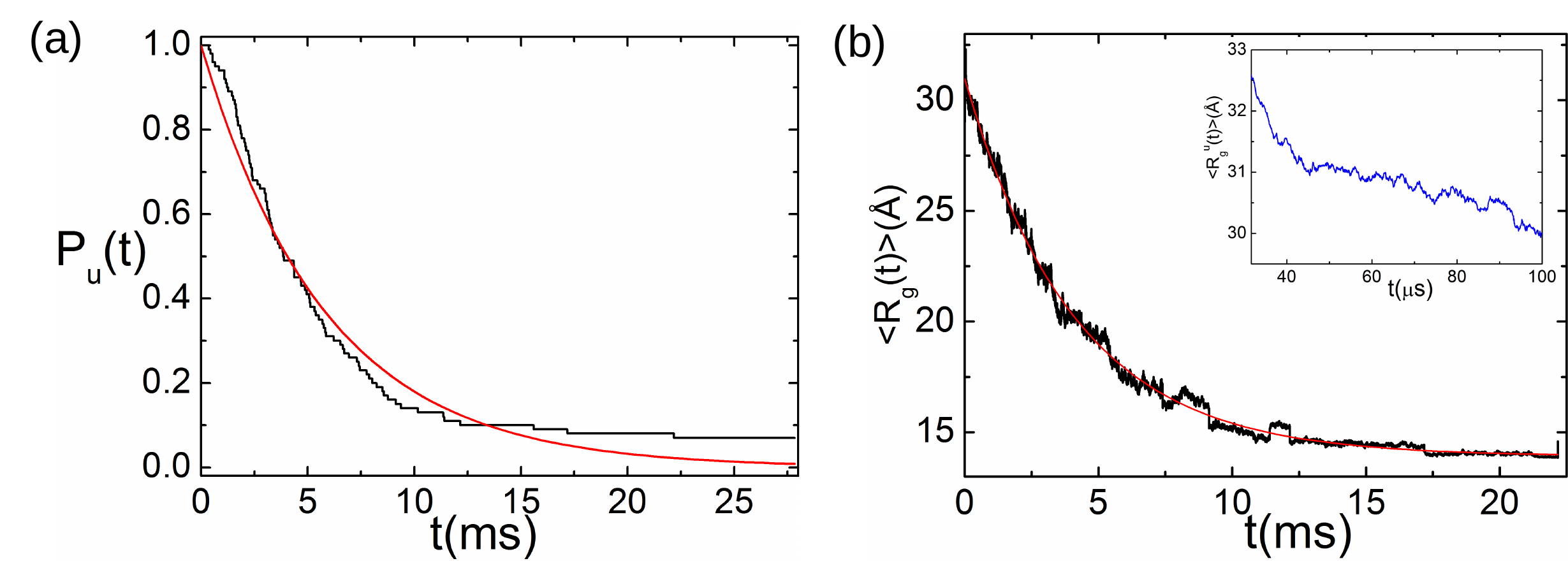}
\caption{}
\end{figure}
\clearpage

\begin{figure}[ht]
\includegraphics[width=\columnwidth]{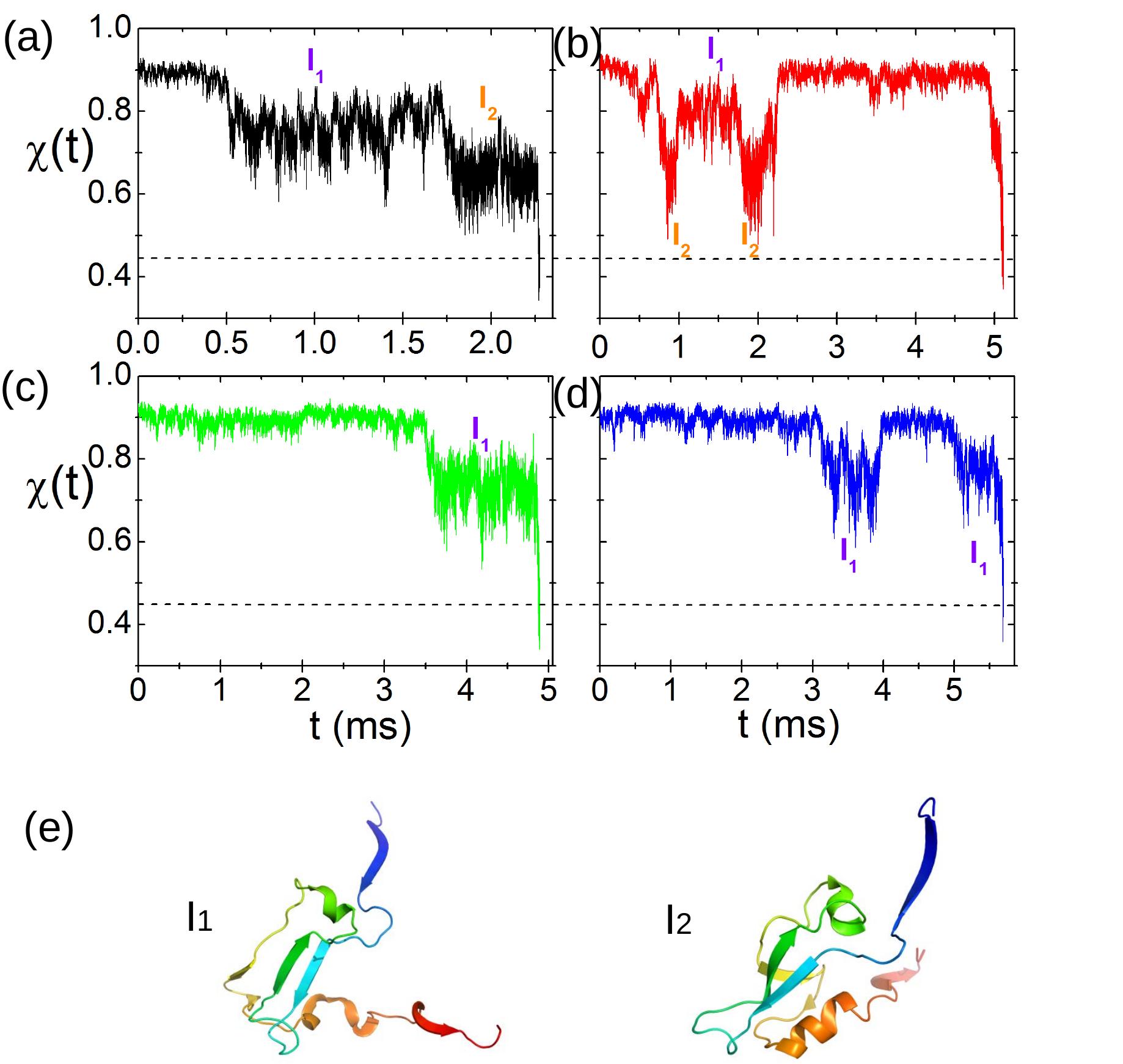}
\caption{}
\end{figure}
\clearpage

\begin{figure}[ht]
\includegraphics[width=\columnwidth]{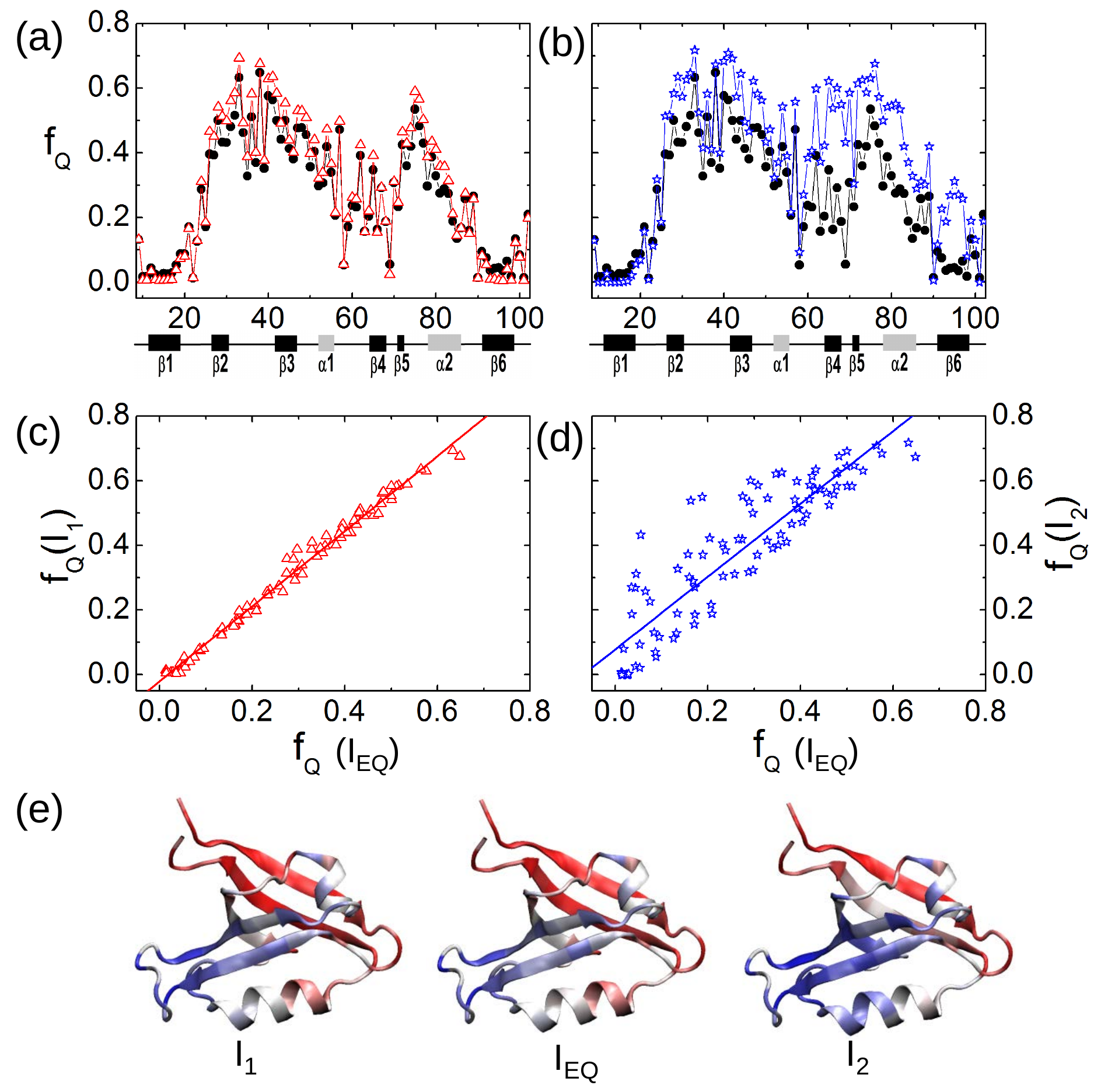}
\caption{}
\end{figure}
\clearpage

\begin{figure}[ht]
\includegraphics[width=\columnwidth]{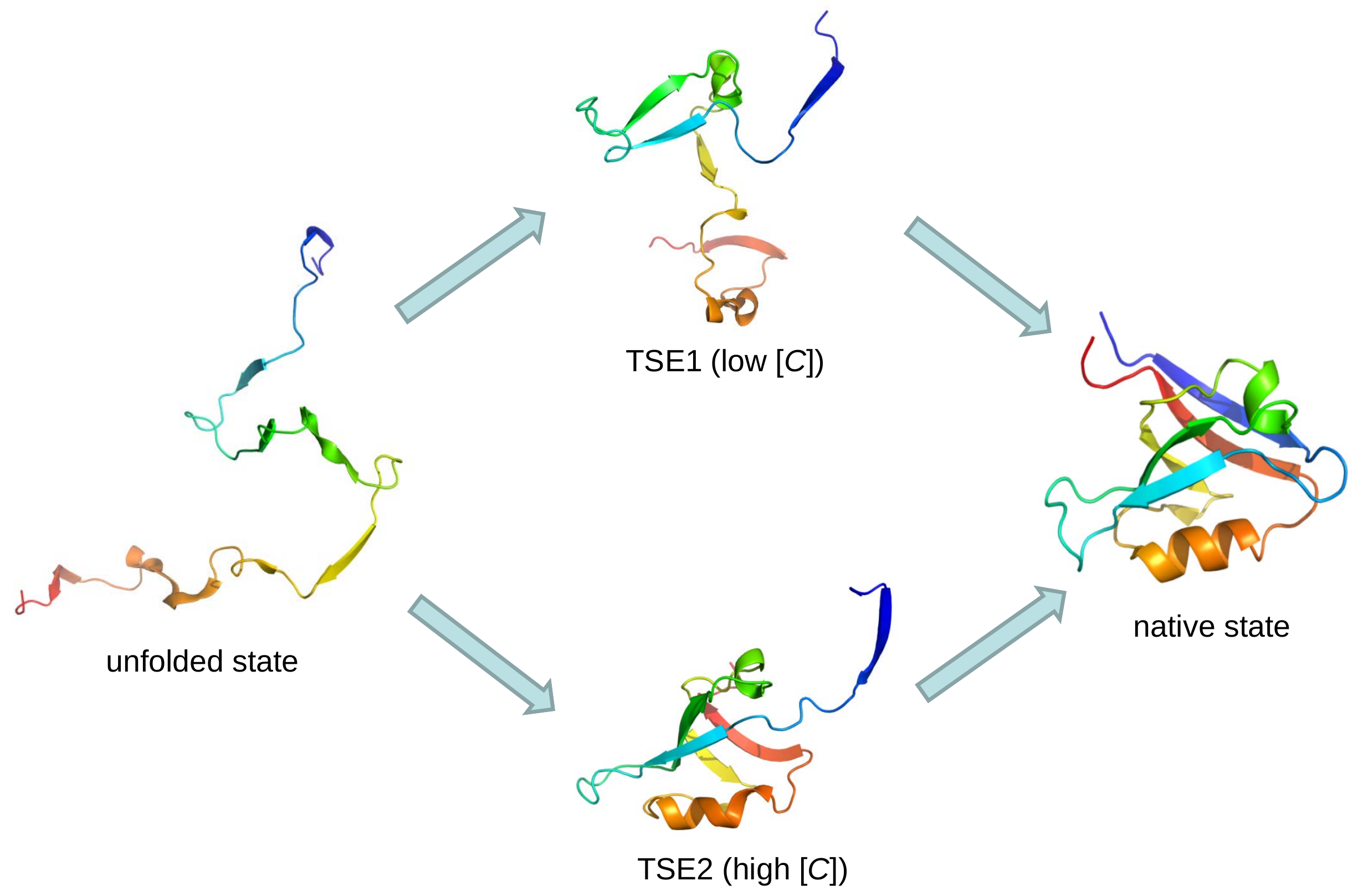}
\caption{}
\end{figure}



\end{document}